\documentclass[useAMS,usenatbib, usegraphicx]{mn2e}
\usepackage{pslatex}

\newcommand{\kms}{\mbox{km}\,\mbox{s}^{-1}}
\newcommand{\cd}{c\,d$^{-1}$}
\newcommand{\target}{V455 And}

\title[Spectral variability in V455 And]{Remarkable spectral variability on the spin period of the accreting white dwarf in \target}

\author[S. Bloemen et al.]{S. Bloemen,$^{1}$\thanks{E-mail:
steven.bloemen@ster.kuleuven.be} D. Steeghs,$^{2,3}$  K. De Smedt,$^{1}$ J. Vos,$^{1}$  B. T. G\"ansicke,$^{2}$ \newauthor T. R. Marsh$^{2}$  and P. Rodriguez-Gil$^{4}$\\
$^{1}$Instituut voor Sterrenkunde, Katholieke Universiteit Leuven, Celestijnenlaan 200 D, B-3001 Leuven, Belgium\\
$^{2}$Department of Physics, University of Warwick, Coventry CV4 7AL, UK\\
$^{3}$Harvard-Smithsonian Center for Astrophysics, 60 Garden Street, Cambridge, MA 02138, USA\\
$^{4}$Departamento de Astrof\'\i sica, Universidad de La Laguna, La Laguna, E-38205, Santa Cruz de Tenerife, Spain\\}

\begin{document}

\date{Accepted xxxx xxx xx. Received xxxx xxx xx; in original form xxxx xxx xx}

\pagerange{\pageref{firstpage}--\pageref{lastpage}} \pubyear{2010}

\maketitle

\label{firstpage}

\begin{abstract}
We present spin-resolved spectroscopy of the accreting white dwarf binary \target. With a suggested spin period of only 67s, it has one of the fastest spinning white dwarfs known. To study the spectral variability on the spin period of the white dwarf, we observed \target\ with 2\,s integration times, which is significantly shorter than the spin rate of the white dwarf. To achieve this cadence, we used the blue arm of the ISIS spectrograph at the 4.2-m William Herschel Telescope, equipped with an electron multiplying CCD (EMCCD). Strong coherent signals were detected in our time series, which lead to a robust determination of the spin period of the white dwarf ($P_{\rm spin}=67.619\pm0.002\,$s). Folding the spectra on the white dwarf spin period uncovered very complex emission line variations in H$\gamma$, He I $\lambda4472$ and He II $\lambda 4686$. We attribute the observed spin phase dependence of the emission line shape to the presence of magnetically controlled accretion onto the white dwarf via accretion curtains, consistent with an intermediate polar type system. We are, however, not aware of any specific model that can quantitatively explain the complex velocity variations we detect in our observations. The orbital variations in the spectral lines indicate that the accretion disc of \target\ is rather structureless, contrary to the disc of the prototype of the intermediate polars, DQ Her. This work demonstrates the potential of electron multiplying CCDs to observe faint targets at high cadence, as readout noise would make such a study impossible with conventional CCDs.
\end{abstract}


%%%%%%%%%%%%%%%%%%%%%%%%%%%%%%%%%%%%%%%%%%%%

\section[]{Introduction}\label{sec_intro}
\target, also known as HS\,2331+3905, is a grazingly eclipsing short period cataclysmic variable ($P_{\rm orb}=81.08\,$mins) identified from the Hamburg Quasar Survey whose variability is very complex \citep{Araujo-BetancorGansicke2005, Pyrzas2011}. It contains a rapidly spinning white dwarf accreting from a low mass companion star. \citet{Araujo-BetancorGansicke2005} found two closely spaced signals at 67.2\,s and 67.6\,s, and labelled the strongest of the two, at 67.2\,s, as the white dwarf spin period. By studying the stability of both signals in a longer photometric dataset spanning 25\,d, \citet{Gansicke2007} concluded that the signal at 67.6\,s must be related to the true spin period. The shorter period signal is more likely to be a beat between the spin period and the $\sim 3.5\,$h periodicity detected in spectroscopy at some epochs \citep{Araujo-BetancorGansicke2005,TovmassianZharikov2007}.  \target\ also shows variability which could be interpreted as white dwarf pulsations.  The system exhibited its first known super-outburst in 2007 \citep{MatsuiUemura2009}. The temperature of the white dwarf is $10\,500\pm500\,$K in quiescence \citep{Araujo-BetancorGansicke2005}. Dwarf nova outbursts heat up the white dwarf, which make \target\ an important test bed for our understanding of the white dwarf instability strip \citep{SzkodyMukadam2010, SilvestriSzkody2012}.

In this study, we focus on fast spectroscopy resolving the spin cycle of the accreting white dwarf. A strong modulation on the white dwarf spin is expected when the white dwarf is sufficiently magnetic to control the accretion flow onto its surface, funnelling the material onto its poles. Cataclysmic variables with a magnetic white dwarf whose magnetic field is not strong enough to synchronise the white dwarf spin with the orbit of the system are known as intermediate polars (IPs). In most of the known IPs, including \target, the region where the accretion flow is controlled by the magnetic field is limited in radial extent from the white dwarf and the bulk of the accretion flow still proceeds via a traditional accretion disc that is steadily fed by the mass donor star.
Studies of the spectral line variations of IPs on the spin period have been performed for a series of IPs with slowly spinning WDs (see e.g., \citealt{HellierMason1987} for a spin trail of Ex Hya, \citealt{HellierMason1990} for FO Aqr, \citealt{HellierCropper1991} for AO Psc, \citealt{Hellier1997} for BG CMi and PQ Gem, and  \citealt{Hellier1999} for V2400 Oph and V1025 Cen). Of the IPs with faster spinning WDs, however, only spin variations of AE Aqr \citep{WelshHorne1998,ReinschBeuermann1994} and DQ Her (\citealt{BloemenMarsh2010_DQHer,MartellHorne1995}) have been studied. The observed changes in the DQ Her spectra remain largely unexplained. No significant spectral line variations have been detected on the spin period of AE Aqr. This could be explained by the fact that the system is probably a propeller system \citep{EracleousHorne1996,WynnKing1997} in which most of the material lost from the companion gets expelled from the system rather than being accreted on the white dwarf.

The earlier papers on IPs \citep[see][for a review]{Patterson1994}, suggested variations in the light curve on the spin period to arise from X-ray reprocessing in the disc or bright spot. Reprocessing of X-rays from the magnetic poles of the WD would lead to sinusoidally shaped brightness enhancements in the spin trails, as the beam sweaps around and illuminates different parts of the disc at different spin phases. More recently, it has become clear that magnetically controlled accretion via accretion curtains is likely to have an important influence on the observed light, by its changing orientation (and hence visibility) during a spin cycle, and possibly by acting as an extra reprocessing region. The idea of curtain-like accretion flows from the disc to the magnetic poles of the white dwarf was originally proposed by \citet{RosenMason1988}.

To better understand the accretion geometry near rapidly spinning white dwarfs, we targeted \target\, which, given its expected spin period of only $\sim67$s, is one of the most rapidly spinning white dwarfs known after AE Aqr \citep[P$_{\rm spin}=33.1$\,s,][]{Patterson1979a,de-JagerMeintjes1994} and V842 Cen \citep[P$_{\rm spin}=56.8\,$s,][]{WoudtWarner2009}. The required observations are technically challenging given the need to obtain relatively high spectral resolution spectroscopy with very short exposure times. We describe the instrumental setup involving an electron multiplying CCD in Section \ref{sec_obs}, detail our custom spectral extraction procedure in Section \ref{sec_red}, followed by our data analysis and interpretation.

%%%%%%%%%%%%%%%%%%%%%%%%%%%%%%%%%%%%%%%%%%%%
%%  OBSERVATIONS
%%%%%%%%%%%%%%%%%%%%%%%%%%%%%%%%%%%%%%%%%%%%

\section[]{Observations} \label{sec_obs}
\target\ was observed with the double armed ISIS spectrograph mounted on the 4.2-m William Herschel Telescope on La Palma (Spain), on the nights of 11, 12 and 13 July 2008. 
Since the integration time can only be of the order of a few seconds, the signal of \target\ would be buried under the readout noise of a conventional CCD. The camera we used, QUCAM2, is equipped with a 1k $\times$ 1k Electron-Multiplying CCD (EMCCD) in which the signal is amplified before readout, such that even the signal of a single photon dwarfs the readout noise. In addition, the CCD has a frame transfer buffer which allows on chip storage of an image such that an exposure can be started while the previous is still being read out. This makes it possible to observe time series of spectra with a negligible dead time of about 12ms between two frames. EMCCDs are more efficient than conventional CCDs when the number of counts per pixel is smaller than the readout noise squared \citep{Marsh2008}, which makes them particulary useful to observe faint targets at high cadence. They are therefore also known as Low Light Level CCDs (L3CCDs). QUCAM2 was used earlier on the red arm of ISIS by \citet{TullochRodriguez-Gil2009}. More information on QUCAM2 can be found in \citet{TullochDhillon2010}.
We paired QUCAM2 with the R1200B grating and a 1" wide longslit to cover 4300 to 4700\,\AA\ at a spectral resolution of 1.5\,\AA. In total 15\,796 blue arm spectra were taken with an integration time of 2.0\,s. The observations are summarized in Table \ref{tab_obs}. 
The slit was held at a fixed sky-angle to permit the simultaneous observation of a brighter comparison star, allowing us to monitor slit-losses and telescope tracking.
%The red arm data were taken with a standard CCD and will be presented elsewhere.

\begin{table*}
 \centering
 \begin{minipage}{170mm}
  \caption{Overview of our spin-resolved spectroscopic observations of \target, performed with an EMCCD at the William Herschel Telescope.}
  \begin{tabular}{@{}lllll lcr@{}}
  \hline
      Date & UT & Conditions & Seeing & Instrument & Grating & Exp. time & Number\\
      \hline
      11-12/07/2008 & 02:02 - 03:59 &  clear & $0.8$ to $1.8$ arcsec & QUCAM2 at ISIS blue arm & R1200B & 2.0s & 3102 \\
      12-13/07/2008 & 01:13 - 05:39 & clear & $\approx 0.7$ arcsec &  & & 2.0s & 7020 \\
      13-14/07/2008 & 01:46 - 05:25 & clear & $0.5$ to $0.8$ arcsec  &  & & 2.0s & 5674 \\
\hline
\label{tab_obs}
\end{tabular}
\end{minipage}
\end{table*}

%%%%%%%%%%%%%%%%%%%%%%%%%%%%%%%%%%%%%%%%%%%%
%%  DATA REDUCTION
%%%%%%%%%%%%%%%%%%%%%%%%%%%%%%%%%%%%%%%%%%%%

\section[]{Data reduction} \label{sec_red}
\texttt{STARLINK}\footnote{\texttt{STARLINK} is open source software and can be obtained from http://starlink.jach.hawaii.edu/starlink .} routines and the \texttt{MOLLY} \footnote{\texttt{MOLLY} and \texttt{DOPPLER} are available for download at http://deneb.astro.warwick.ac.uk/phsaap/software/ .}  \newcounter{footnotesoftware}\setcounter{footnotesoftware}{\value{footnote}} software packages written by TRM were used for the data reduction. 

% -------------------------------------------------------------------
%  Gain
% -------------------------------------------------------------------

\subsection{Gain determination}
\begin{figure}
\includegraphics[width=84mm]{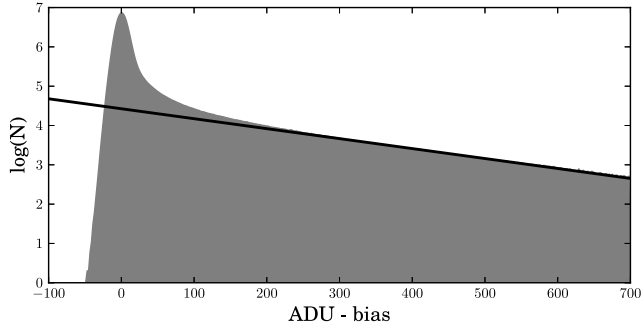}
 \caption{Histogram with 400 bins of the pixels values around the bias level of the pixels in the overscan region of the science spectra. All pixels are expected to have had a charge of 0 or 1 electron. The Gaussian distribution centred at 0 represents the pixels with a zero charge. The pixels with a charge of one electron give rise to the tail, that extends to very hight count values. The slope of the linear fit to the tail is $-1/g$. }
  \label{FIG_gain}
\end{figure}

The gain of an EMCCD is highly dependent upon the voltage that is applied to the serial register that amplifies the signal before readout. We determined the gain of the EMCCD from the overscan regions of the science spectra. Since these have no illumination, the number of electrons is zero in most pixels except in a few pixels because of clock induced charges. The probability distribution for the output of 1 input electron is approximately
\[p(x)_{n=1}=\frac{\exp{\left(-x/g\right)}}{g}\]
(\citealt{BasdenHaniff2003}). The histogram in Fig.~\ref{FIG_gain} shows the logarithm of the distribution of the pixel values of the overscan regions of all science frames. The Gaussian distribution around 0 results from the pixels with zero charge. The linear part results from the pixels with a single electron charge. The number of pixels with a higher charge is negligible. The slope of the linear part is $-1/g$, from which we derive $g=170 \pm 8$ for the combined data of the three nights. There was no evidence for significant gain variations between different nights, thus we use the average gain for all extractions. This value is slightly higher than the nominal gain of 160 mentioned in the instrument manual, but longer-term variations at this level are to be expected.

% -------------------------------------------------------------------
%  Extraction
% -------------------------------------------------------------------

\subsection{Extraction and calibration of curved spectra}
The spectra of \target\ were slightly curved and tilted. Therefore, the spectra were extracted along a 3rd order polynomial fit to the spectrum (trace) following \citet{Marsh1989}. Because the advantage of EMCCDs is largest when working at low count levels \citep{Marsh2008}, the science spectra typically have low count levels. This prevents one from accurately fitting the trace of the curved spectra. If the position of the spectra on the CCD is constant in time, this problem can be overcome by stacking lots of frames.

Unfortunately, the positions of the object spectra were found to be wobbling along the spatial axis on timescales of roughly a minute with a semi-amplitude of 2 to 3 pixels, equivalent to about 0.5 arcsec. 
To deal with this extra complication, we implemented a multistep reduction strategy which replaces the standard reduction techniques \citet{TullochRodriguez-Gil2009} used to reduce QUCAM2 data. First, we determined a 3rd order polynomial trace and a profile for optimal extraction for both stars from a stacked image. Then, we tweaked the 
first order parameter of the trace of the (brighter) comparison star on each frame. Subsequently, we shifted the frames to align the spectra of the comparison star, and thus also the spectra of the target assuming a constant offset between the two on short timescales. From the shifted frames, a new stack was made which was used to recompute the traces and profiles. These traces and profiles were used to extract the spectra of the two stars from the original frames, tweaking the first order parameter of the trace of the comparison star and applying the same shift to the trace of the target star. The sky regions that were used for the sky subtraction were shifted by the same number of pixels.

Arc spectra were taken every $\sim 2$ hours and wavelength calibrated by a third order polynomial fit to the lines achieving an RMS of $\sim0.05\,$\AA. The wavelength scale of the \target\ spectra was determined by linear interpolation between bracketing arc spectra. Spectra of HZ44 were taken every night and used to flux-calibrate the \target\ spectra. Slit losses of individual exposures were accounted for by comparison of the flux level of the spectrum of the in-slit comparison star with a wide slit spectrum.

%%%%%%%%%%%%%%%%%%%%%%%%%%%%%%%%%%%%%%%%%%%%
%%  Analysis
%%%%%%%%%%%%%%%%%%%%%%%%%%%%%%%%%%%%%%%%%%%%

\section[]{Analysis} \label{sec_analysis}

\begin{figure}
\includegraphics[width=84mm]{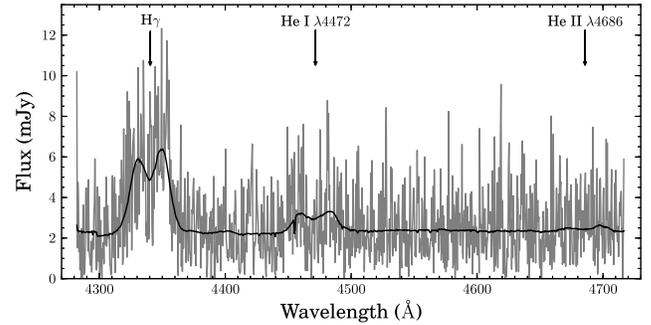}
 \caption{ The average (black) blue arm spectrum of \target\ overplotted on a single spectrum (grey).}
  \label{FIG_av_blue}
\end{figure}

Fig.~\ref{FIG_av_blue} shows an example single exposure (grey) and the average (black) spectrum of \target. Double peaked line profiles of H$\gamma$ $\lambda 4340$ and He I $\lambda 4471$, characteristic for accretion disc emission, can clearly be seen. He II $\lambda 4686$ is visible as well, but is much weaker.

% -------------------------------------------------------------------
%  Frequency analysis
% -------------------------------------------------------------------

\subsection{Frequency analysis}

\begin{figure}
\includegraphics[width=84mm]{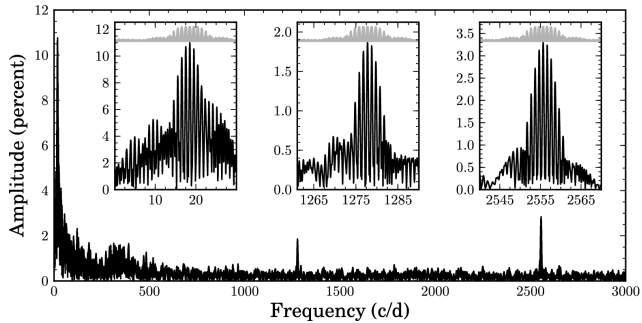}
 \caption{Scargle periodogram of the flux in H$\gamma$ above the continuum. In the insets, the window function is plotted in grey, shifted to the frequency of the strongest peak. The highest peak in the left inset is 18.5\,\cd\, which is the one day alias of the orbital frequency peak at 17.5\,\cd, which is almost equally strong. In the second inset, the highest amplitude is found at 1277.8\,\cd\ which is identified as the spin period of the white dwarf. The peak in the third inset, at 2555.6\,\cd, is the first harmonic of this spin period.}
  \label{FIG_scargle_4340}
\end{figure}

To search for oscillations in the emission lines, we present a Lomb-Scargle periodogram \citep{Scargle1982, Lomb1976} of the net line flux in H$\gamma$ in Fig.~\ref{FIG_scargle_4340}. The net line flux was obtained by integrating the flux in the line (above the continuum level) over a region of 60\,\AA\ centered at H$\gamma$. Our short exposure times allow us to resolve line flux modulations out to several thousand cycles/day. The highest peaks are found at the orbital frequency of 17.5\,\cd, and its one-day aliases (the window function is shown in grey in the insets, centered at the highest peak). 
In addition, the highest amplitude peak in the second inset is at $1277.8$ \cd. This coherent signal would be most naturally 
identified as the spin period of the white dwarf, as IPs often show a spin-locked signal in their emission lines. Furthermore, no peak is found above noise level at 1284.7\,\cd, which \citet{Araujo-BetancorGansicke2005} suggested to be WD spin frequency but is more likely a beat frequency, as explained in Section \ref{sec_intro}. The third inset  zooms in on the region around 2555.6\,\cd, the first harmonic of the spin period whose peak power is higher than the fundamental.
The variation of the spectra on the orbital period is further discussed in Section \ref{sec_orbit} and the variability on the spin period in Section \ref{sec_spin}.

We do see excess power in the emission line power spectrum near the nominal pulsation period range ($\sim 300\,$\cd). This implies that the pulsation source is not a pure continuum source, but must have a variability component in the H$\gamma$ line. This could be caused by variability in the underlying Balmer absorption profile from a pulsating white dwarf, but is also consistent with a variable contribution from the line emission sources. Our data cannot distinguish between these two given that both the white dwarf absorption line and the accretion powered emission line have comparable widths and thus, at all velocities where we detect variability, overlap \citep[see Fig.\ 12 in ][]{Araujo-BetancorGansicke2005}.
 
 We have also checked the variability in the continuum part of the spectra (between 4520 and 4670\,\AA). We found variability at the same frequencies as in the H$\gamma$ line, but at lower amplitude, except for the variability at 1277.8\,\cd which is not detectable above the noise level. This is in agreement with the photometric studies mentioned earlier, which find a much lower variability amplitude at 1277.8\,\cd than at the first harmonic of this frequency.

% -------------------------------------------------------------------
%  Orbital trails
% -------------------------------------------------------------------

\subsection{Variations on the orbital period}\label{sec_orbit}

\begin{figure*}
\includegraphics[width=180mm]{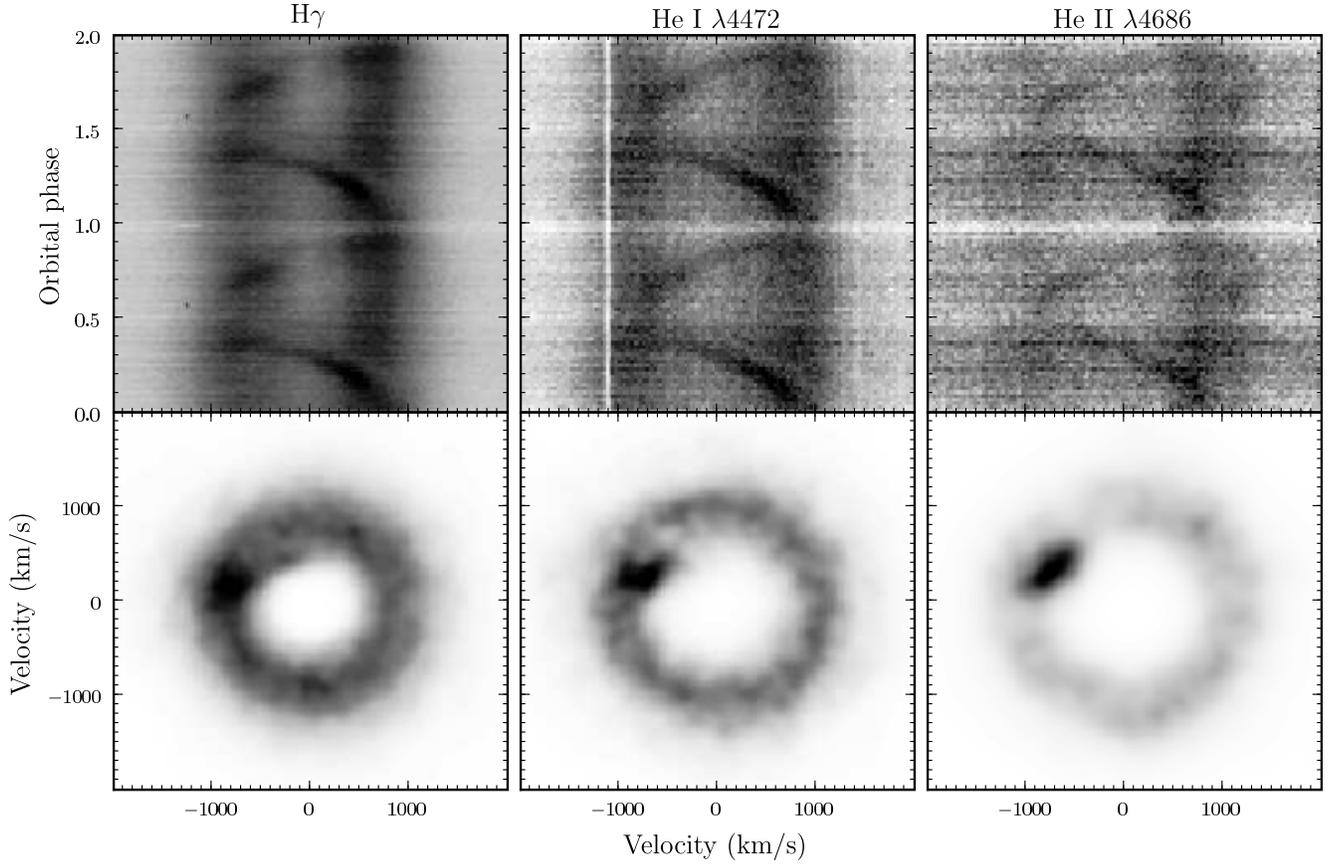}
 \caption{Orbital trails and Doppler maps for H$\gamma$, He I $\lambda4472$, He II $\lambda 4686$. White is minimum flux, black maximum flux. A grazing eclipse can be seen at orbital phase 1. The three lines result in comparable Doppler maps that show emission of the accretion disc (ring) and the bright spot where the mass stream from the companion hits the disc (brighter dot at the ring).}
  \label{FIG_orbitaltrails}
\end{figure*}

We rebinned the spectra on a constant velocity scale of $28\,\kms$ per pixel and folded the spectra on the orbital ephemeris of \citet{Araujo-BetancorGansicke2005}. The resulting orbital trails for H$\gamma$, He I $\lambda4472$ and He II $\lambda4686$ are shown in the upper panels of Fig.~\ref{FIG_orbitaltrails}. Apart from the double peaked profile emitted by the accretion disc, a strong bright spot contribution is seen as an S-wave in the trails. A weak WZ Sge-like `bright spot shadow' \citep[][]{SpruitRutten1998} can be seen in the red-shifted part of the lines at orbital phase $\sim 0.25$. Around orbital phase 1, a shallow eclipse is observed. In all three lines, the red-shifted peak appears to be brighter than the blue-shifted peak at all orbital phases. This difference is probably due to a slightly warped or slowly precessing accretion disc of which the red-shifted surface was better visible and/or brighter at the time of our observations compared to the blue-shifted side. Comparing night-by-night averages reveals no significant evolution in this line asymmetry across our three nights.

The bottom panels of Fig.~\ref{FIG_orbitaltrails} show the Doppler maps of the trails, produced using the maximum entropy method as presented by \citet{MarshHorne1988} and  implemented in the \texttt{DOPPLER}\footnotemark[\value{footnotesoftware}] package. The three maps are not significantly different. The accretion disc emission maps onto a diffuse ring. The disc appears to be structureless, as is the case for most IPs but contrary to the prototype DQ Her, which was found to have spiral density structures \citep{BloemenMarsh2010_DQHer}. The dot around $($V$_{X}$,V$_{Y}$$)\approx(-900,100)\,\kms$ represents the strong brightspot emission component which is hot enough to contribute significantly to HeII. No line emission from either the white dwarf or the secondary star is detected. These maps show that the bulk of the line emission originates from an extended accretion disc, implying that any magnetically controlled flow is confined to the zone near the white dwarf.

% -------------------------------------------------------------------
%  Spin trails
% -------------------------------------------------------------------

\subsection{Variations on the WD spin period}\label{sec_spin}

\begin{figure*}
\includegraphics[width=180mm]{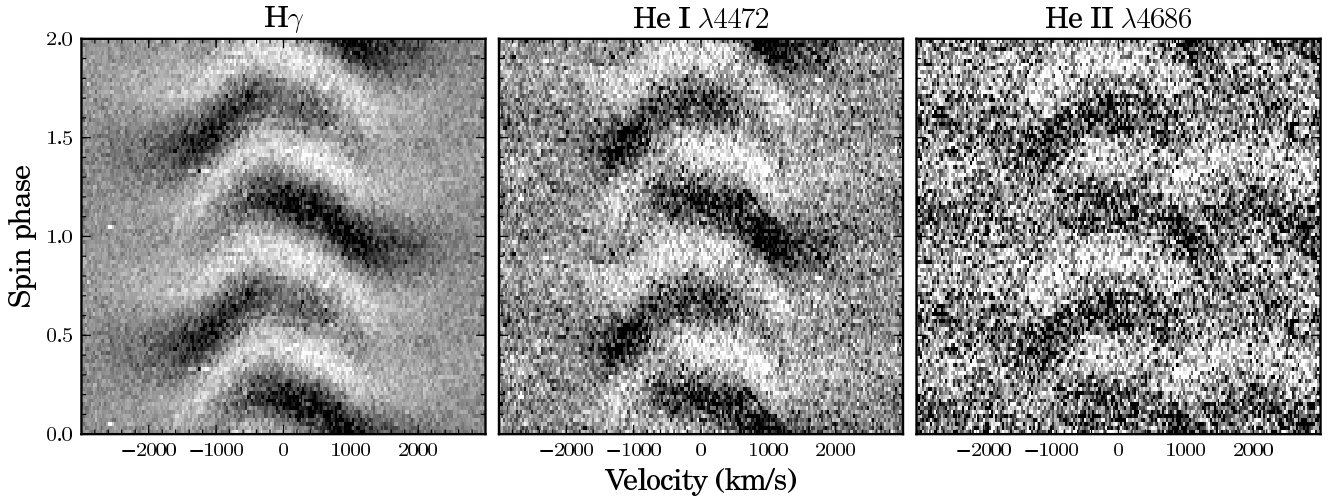}
 \caption{Trails of normalised and mean subtracted line profiles, folded on \target's spin period of 67.619\,s. The color scales are $\pm$ 0.5, 0.2 and 0.1 mJy compared to average. White indicates lower than average fluxes, black higher than average.}
  \label{FIG_spintrails}
\end{figure*}

To exploit our EMCCD setup and study the changes in the line profiles at the white dwarf's spin period, the spectra were phase folded into 50 bins at the spin period of $67.619$\,s (1277.75\,\cd). An average spectrum was subtracted to highlight the spin modulation, similar to the procedure outlined in \citet{BloemenMarsh2010_DQHer}. The resulting spin trails are shown in Fig.~\ref{FIG_spintrails} for the three spectral lines. The detection of a clear pattern is strong evidence for the fact that, as concluded by \citet{Gansicke2007}, 67.619\,s is the true spin period because folding the spectra on slightly different periods, such as the 67.24\,s period that was suggested by \citet{Araujo-BetancorGansicke2005}, washes out the structure. In DQ Her there is still a debate whether the spin signal is at the spin period or its harmonic. Here we can confidently conclude that the folded trails robustly designate $67.619\pm0.002$\,s to be the rotation period of the white dwarf, with the power detected at the 1st harmonic caused by the two-sided pattern in Fig.~\ref{FIG_spintrails}. The uncertainty on the spin period was determined by comparing how well spin trails of the different nights phase up, after folding the data on various different periods that are slightly different from the optimal spin period value. 

\begin{figure}
\includegraphics[width=84mm]{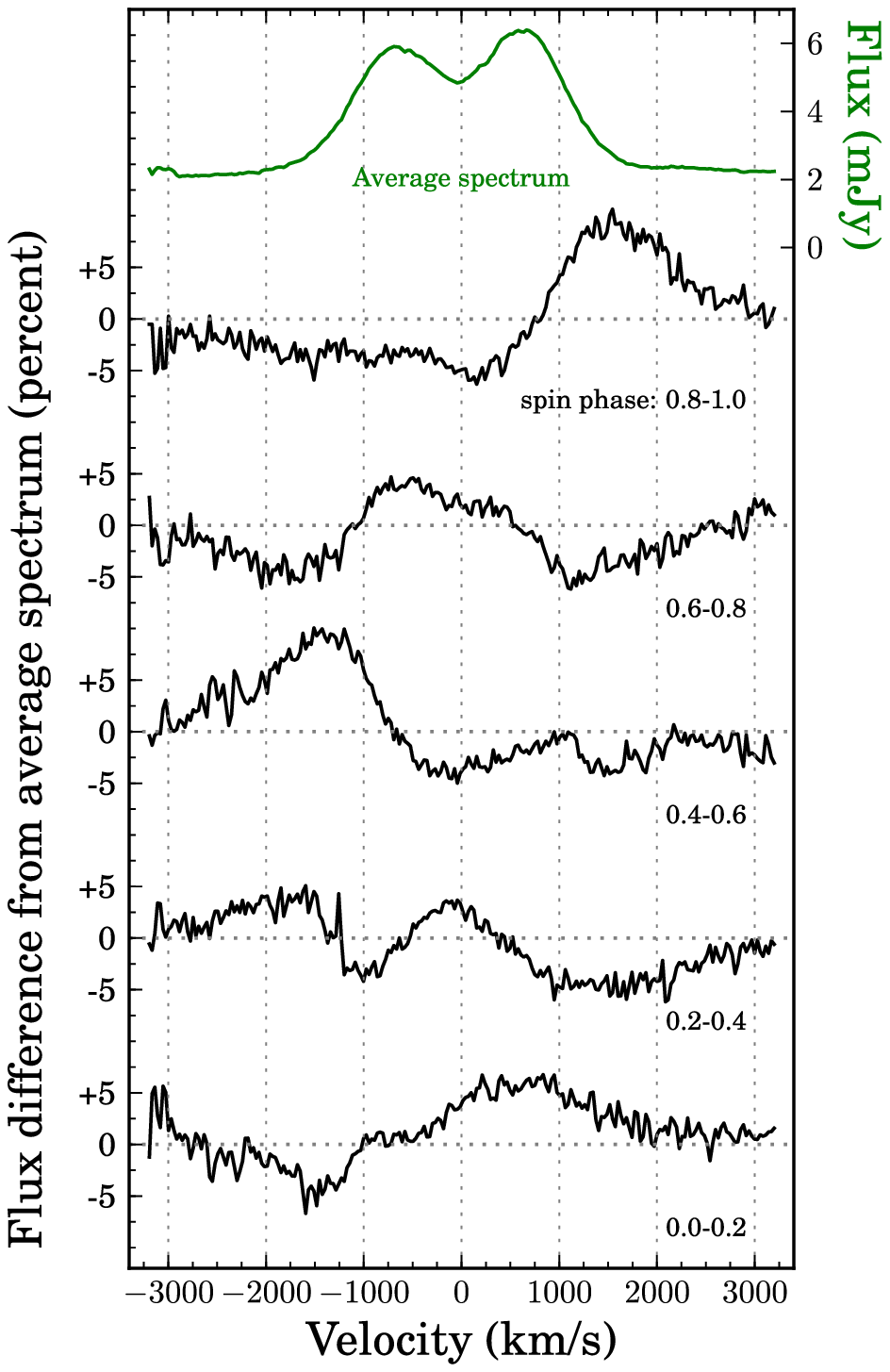}
 \caption{Five spin profiles for the H$\gamma$ line normalised to show the spin modulation in percent compared to the average line profile at the top. The spin modulation peaks at velocities well beyond the disc peaks and can be detected out to large radial velocities.  }
  \label{FIG_spinprofiles}
\end{figure}

The observed pattern is puzzling. Previously published spin trails of IPs with relatively slowly spinning IPs (see references in Section \ref{sec_intro}) appear to be totally different to the trails we find for \target. Also the recently published spin trails of DQ Her \citep[see][]{BloemenMarsh2010_DQHer}, although also being an IP with a fast-spinning WD, are fundamentally different. Firstly, DQ Her's variations in He II $\lambda4686$ are found to be much stronger in the red-shifted parts of the line than in the blue-shifted, which is clearly not the case for \target. 
The spin modulations can be detected out to large velocities and their amplitudes peak at velocities significantly beyond the disc peak. To illustrate this, we plot several representative normalised spin profiles in Fig.~\ref{FIG_spinprofiles}. These illustrate the amplitude of the observed spin signal as a function of velocity, with an average emission line profile provided as a reference.
The structure in the spin trials and profiles cannot be explained by the `old school' hypothesis of X-ray reprocessing in the disc and the brightspot. 
If reprocessing of X-rays by the bright spot contributed significantly, we would expect a higher spin modulation amplitude in the spectra lines at the radial velocity of the bright spot, which would cause the spin pattern to be orbital phase dependent. Spin trails were therefore also produced using subsets of the spectra, taken at selected orbital phase intervals, but no variability of the spin trails over the orbital period could be detected.
The complex pattern more likely results from magnetically controlled accretion near the white dwarf. Accretion curtains flowing towards the white dwarf would involve large velocities, switching from maximum to minimal radial velocity as they rotate along the line of sight across 1/4 spin cycle. However, we do not know which exact geometry can explain the observed modulations. A thorough modelling effort of the accretion curtain geometry and its optical properties will be required to quantitatively test this idea and see whether such a geometry could reproduce the observed spin modulations as a function of radial velocity and spin phase. 

%%%%%%%%%%%%%%%%%%%%%%%%%%%%%%%%%%%%%%%%%%%%
%%  CONCLUSIONS
%%%%%%%%%%%%%%%%%%%%%%%%%%%%%%%%%%%%%%%%%%%%

\section[]{Summary} \label{sec_concl}
We observed the intermediate polar \target\ with the electron multiplying CCD `QUCAM2' installed at the ISIS spectrograph at the William Herschel Telescope. We demonstrate the potential of this detector to observe faint targets at high cadence. With a conventional detector, spin-resolved spectroscopy of the IPs with the fastest spinning WDs would be hard if not impossible to achieve as readout noise would swamp any signal. We developed a strategy to reduce the faint spectra, making full use of a brighter in-slit comparison star and correcting for instrument and telescope flexure. 
We were able to detect strong coherent signals in our time-series allowing us to robustly identify the spin period of the white dwarf to be  $67.619\pm0.002$\,s, which confirms the spin period reported by \citet{Gansicke2007}. Furthermore, by folding our 15\,796 spectra on this spin period, a complex emission line variation can be recovered resolving the spin modulation as a function of both radial velocity as well as spin phase.
The observed variations are totally different from the results of previous observations of other intermediate polars, including the canonical IP DQ Her. We believe that the observed patterns are evidence of magnetically controlled accretion curtains near the white dwarf, but are not aware of any specific model that can reproduce our observations in detail.

% -------------------------------------------------------------------
%  Acknowledgements
% -------------------------------------------------------------------

\subsection*{Acknowledgements}
The observations were made with the William Herschel Telescope operated at the island of La Palma by the Isaac Newton Group in the Spanish Observatorio del Roque de los Muchachos of the Instituto de Astrof\'isica de Canarias. We would like to thank the ING staff for preparing and supporting the use of the QUCAM2 detector.

The research leading to these results has received funding from the
European Research Council under the European Community's Seventh Framework Programme
(FP7/2007--2013)/ERC grant agreement n$^\circ$227224 (PROSPERITY).
During this research DS and TRM were supported under grants from the UK's 
Science and Technology Facilities Council (STFC, ST/F002599/1 and 
PP/D005914/1 Advanced Fellowship).

%%%%%%%%%%%%%%%%%%%%%%%%%%%%%%%%%%%%%%%%%%%%
%%  BIBLIOGRAPHY
%%%%%%%%%%%%%%%%%%%%%%%%%%%%%%%%%%%%%%%%%%%%

\bibliographystyle{mn2e}
\bibliography{bibtex}

\label{lastpage}

\end{document}